\documentclass[lettersize,journal]{IEEEtran}
\usepackage{amsmath,amsfonts}
\usepackage{algorithmic}
\usepackage{algorithm}
\usepackage{array}
\usepackage[caption=false,font=normalsize,labelfont=sf,textfont=sf]{subfig}
\usepackage{textcomp}
\usepackage{stfloats}
\usepackage{url}
\usepackage{verbatim}
\usepackage{graphicx}
\usepackage{cite}
\usepackage{booktabs}
\usepackage{multirow}
\usepackage{pifont}
\usepackage{hyperref}
\begin{document} 

\title{Mixture of LoRA Experts with Multi-Modal and Multi-Granularity LLM Generative Error Correction for Accented Speech Recognition}

\author{Bingshen Mu, Kun Wei, Pengcheng Guo, Lei Xie,~\IEEEmembership{Senior Member,~IEEE}
\thanks{Corresponding author: Lei Xie.}
\thanks{Bingshen Mu, Kun Wei, Pengcheng Guo, and Lei Xie are with Audio, Speech and Language Processing Group (ASLP@NPU), School of Computer Science, Northwestern Polytechnical University, Xi’an 710072, China (e-mail: bsmu@mail.nwpu.edu.cn; guopengcheng1220@gmail.com; lxie@nwpu.edu.cn).}
}

\markboth{Journal of \LaTeX\ Class Files,~Vol.~14, No.~8, August~2021}%
{Shell \MakeLowercase{\textit{et al.}}: A Sample Article Using IEEEtran.cls for IEEE Journals}


\maketitle

\begin{abstract}
Despite improvements in automatic speech recognition, performance drops with accented speech.
Generative error correction (GER) leverages the linguistic knowledge of large language models (LLMs), outperforming typical language model methods.
However, it lacks specificity in accented speech scenarios.
Accents represent deviations from standard pronunciation, making multi-granularity pronunciation and semantic information essential for accented speech recognition.
Moreover, accents exhibit considerable diversity, with each accent possessing distinct characteristics.
In this study, we leverage GER to improve transcription accuracy by addressing the two primary features.
We propose the \textit{multi-modal GER}, which integrates pronunciation information from the speech modality, and the \textit{multi-granularity GER}, which incorporates fine-grained phoneme-level pronunciation information.
These methods enable the LLM to utilize the pronunciation information of accented speech and the semantic information from word-level hypotheses for accurate transcription predictions through low-rank adaptation (LoRA) fine-tuning.
We employ a three-stage strategy to train separate multi-modal GER models for each accent to obtain mono-accent LoRA experts.
By adopting our proposed \textit{HDMoLE} method, which incorporates hierarchical routing and dynamic thresholds within the mixture of LoRA experts, we effectively merge mono-accent LoRA experts within a single multi-modal GER to overcome accent diversity challenges.
Furthermore, multi-granularity GER leverages N-best word-level and phoneme-level hypotheses from the HDMoLE model to predict final transcriptions.
Experiments on a multi-accent English dataset show that our methods reduce word error rate by 67.35\% compared to the baseline vanilla Whisper-large-v3 model.
\end{abstract}


\section{Introduction}
\IEEEPARstart{A}{utomatic} speech recognition (ASR) has become an increasingly crucial technology in contemporary society because it can efficiently and precisely transcribe spoken language.
With the continuous advancement of deep learning, ASR has experienced remarkable improvements in its efficacy~\cite{graves2012sequence, bahdanau2016end, chan2016listen, dong2020cif, watanabe2017hybrid}.
Nevertheless, ASR struggles with numerous errors when confronted with speech variations stemming from diverse factors such as background noise~\cite{mu2024automatic} and speaker accents~\cite{mu2024mmger}.
Such detrimental factors are widespread and inevitable in speech signals, significantly affecting the precision of the recognition results.
Humans exhibit extraordinary adaptability to variations in speech environments. Our recognition system surpasses dependence on speech alone, frequently interpreting ambiguous or distorted speech through contextual understanding and innate linguistic knowledge.
Likewise, current ASR models commonly employ independent language models (LMs) for rescoring during decoding~\cite{kannan2018analysis, hu2020deliberation, shan2019component, sriram2017cold}.
Using beam search decoding, ASR models produce N-best hypotheses, which are then scored by independent LMs. The hypotheses with the highest score are selected as the final ASR outputs.
Additionally, LMs are extensively employed in ASR error correction tasks, leveraging either the 1-best or N-best hypotheses generated by ASR models~\cite{leng2021fastcorrect,mani2020asr, leng2023softcorrect, ma2023n}.

The impressive development of large language models (LLMs)~\cite{achiam2023gpt, chowdhery2023palm, anil2023palm, touvron2023llama, touvron2023llama2} has significantly spurred extensive research exploring their potential across various fields, particularly in ASR.
LLMs' advanced contextual reasoning capabilities and comprehensive textual knowledge make them a more promising component than conventional LMs for providing semantic guidance in ASR.
Driven by LLMs, ASR generative error correction (GER)~\cite{chen2024hyporadise} has achieved exceptional performance in learning the mapping from hypotheses to transcriptions through LLM parameter-efficient fine-tuning, significantly outperforming typical LM rescoring and error correction methods.
Nevertheless, GER proves constrained efficacy in generating accurate transcriptions under unfavorable speech environments.
To enhance GER's performance in noise-robust scenarios, RobustGER~\cite{hu2024large} extracts a noise embedding in the language space of N-best hypotheses generated by ASR and applies knowledge distillation to distill the real noise information in speech embeddings into the language space noise embedding.
This process achieves efficient language space denoising and demonstrates notable noise robustness.

However, GER still lacks specificity in accented speech scenarios, which are the most common in the real world.
Accents represent deviations from standard pronunciation norms influenced by the speaker’s educational background, geographical region, or native language~\cite{markl23_interspeech}.
In contrast to noise-robust speech recognition, accented speech recognition demands both pronunciation and semantic information~\cite{shao2023decoupling}.
Different accents exhibit considerable variations in the pronunciation of phonemes, potentially leading to the emergence of phoneme variants. 
Additionally, accents may involve distinctive phenomena such as phoneme linking, reduction, substitution, and elision. 
These nuanced pronunciation differences are essential for achieving precise accented speech recognition.
GER depends exclusively on N-best word-level hypotheses for transcription prediction, omitting pronunciation information inherent in accented speech.
Consequently, incorporating speech embeddings generated by the speech encoder or N-best phoneme-level hypotheses as fine-grained pronunciation information into GER offers a viable solution. This solution enables GER to leverage pronunciation and semantic information for more accurate accented speech transcription prediction.

Moreover, a single language often encompasses various accents in the real world.
As an illustration, English exhibits several prominent accents, such as British, American, and Indian, each distinguished by distinctive pronunciation patterns.
Training a dedicated model for each accent and treating it as an expert, then integrating these experts through the Mixture of Experts (MoE)~\cite{shazeer2017outrageously, fedus2022review, zoph2022st, yuan2023moec, song2024u2++, kwon2023mole} methods, is a practical solution for addressing accent diversity challenges in accented speech recognition.
The MoE is an ensemble method commonly viewed as a collection of experts, each focusing on different domains.
A trainable gating network (router) assigns weights to these experts.
Low-rank adaptation (LoRA)~\cite{hu2021lora} fine-tuning in GER allows the LLM to learn the mapping from hypotheses to transcriptions.  
Similarly, the LoRA method can train the LLM to identify and capture the distinctive features of various accents, thereby enhancing its ability to map hypotheses to transcriptions with precision under particular accent conditions.
Inspired by MoE, numerous researchers regard LoRA as a domain expert to overcome the challenges encountered by large models in real-world multi-domain scenarios, leading to the proposal of the mixture of LoRA experts (MoLE) methods~\cite{liu2023moelora, feng2024mixture, zhu2023sira, li2024mixlora, yang2024moral, dou2024loramoe, mu2024hdmole, wu2024mixture}.
Significant differences may exist among accents, and combining LoRA experts specialized in each accent through MoLE methods is a feasible solution for addressing multi-accent speech recognition challenges.
While the various MoLE methods offer valuable insights into integrating MoE and LoRA, numerous challenges persist.
The high coefficient of variation in unconstrained MoE layers reflects the router's consistent assignment of larger weights to the same few experts~\cite{shazeer2017outrageously}.
This expert utilization imbalance problem is typical in MoE and indicates a lack of clear correspondence between experts and domains.
Moreover, the static Top-K expert selection strategy constrains MoE's adaptability, highlighting the need for more dynamic expert selection strategies to address the complexities of different domains~\cite{liu2024adamole}.

This study explores the application of GER and MoLE to address the speech recognition challenges posed by real-world multi-accent scenarios.
To comprehensively exploit the pronunciation information in accented speech, we propose the \textit{multi-modal GER}, which integrates pronunciation information from the speech modality, and the \textit{multi-granularity GER}, which incorporates fine-grained phoneme-level hypotheses associated with pronunciation.
These two methods empower the LLM to utilize the pronunciation information from accented speech with the semantic information provided by word-level hypotheses for precise transcription predictions through LoRA fine-tuning.
On the one hand, we implement a three-stage training strategy for the multi-modal GER model using various mono-accent speech data, which results in corresponding mono-accent LoRA experts.
Using the MoLE method, all the mono-accent LoRA experts are merged into a single multi-modal GER model, effectively addressing the challenges posed by accent diversity.
Our previous study~\cite{mu2024hdmole} proposed a novel MoLE method with hierarchical routing and dynamic thresholds named \textit{HDMoLE}. 
In contrast to the earlier HDMoLE, which trains LoRA experts and local routers from scratch, this study employs pre-trained mono-accent LoRA experts mentioned above, whose parameters remain fixed.
HDMoLE improves transcription prediction accuracy by selecting optimal combinations of LoRA experts for accented speech without compromising their existing capabilities.
On the other hand, multi-granularity GER leverages N-best word-level and phoneme-level hypotheses generated from the HDMoLE model to achieve an optimal balance between pronunciation precision and semantic understanding for predicting the final accented speech transcriptions.

In general, our proposed methods can be summarized as a pipeline: (1) multi-modal GER to obtain multiple mono-accent LoRA experts; (2) HDMoLE to combine LoRA experts; (3) multi-granularity GER to predict final transcriptions.
In this pipeline, the three methods follow a progressive relationship. 
The main purpose of multi-modal GER is to generate multiple mono-accent LoRA experts. 
These LoRA experts are then combined through HDMoLE to handle multi-accent scenarios and produce preliminary error correction results. 
Subsequently, the multi-granularity GER further refines the error correction based on the preliminary results to generate the final predicted transcription.
Additionally, we briefly outline the behaviors of each method to explain how these methods collectively contribute to solving the target problem:
\begin{itemize}
    \item \textbf{Multi-modal GER:} The pronunciation information from the speech modality in accented speech and the semantic information from the text modality of the ASR 1-best hypotheses are used to fine-tune the LLM through LoRA to accurately predict transcriptions while also obtaining mono-accent LoRA experts.
    \item \textbf{HDMoLE:} Integrate the outputs of mono-accent LoRA experts through hierarchical routing and dynamic thresholds to adapt to multi-accent scenarios.
    \item \textbf{Multi-granularity GER:} The pronunciation information from the N-best phoneme-level hypotheses and the semantic information from the word-level hypotheses generated by HDMoLE are used to fine-tune the LLM through LoRA, enabling it to predict final transcriptions.
\end{itemize}
Through the progressive collaboration of the three methods, the accuracy of the predicted accented speech transcriptions and the generalization across accents are significantly improved.
Experimental results on the multi-accent English dataset demonstrate the efficacy of our proposed methods. Compared to the vanilla Whisper-large-v3 baseline model~\cite{radford2023robust}, our methods achieve a remarkable relative word error rate (WER) reduction of 67.35\%.

\section{Related Works}
This section first provides an overview of language modeling in ASR, including typical LM methods and GER.
Then, we summarize using multi-granularity information in accented speech recognition.
Subsequently, we introduce the LoRA method for parameter-efficient fine-tuning of LLMs within GER, followed by various MoLE methods for combining multiple LoRA experts.
\subsection{Language Modeling in ASR}
LM rescoring has been widely adopted to improve the linguistic acceptability of ASR results, consistently producing stable performance gains across various ASR models.
Generally, an external LM is trained independently and applied to rescore the N-best hypotheses produced during ASR beam search decoding.
Numerous ASR and LM fusion methods have been introduced, including shallow fusion~\cite{chorowski2016towards, zeyer2018improved, toshniwal2018comparison, kannan2018analysis}, component fusion~\cite{shan2019component}, cold fusion~\cite{sriram2017cold}, and deliberation~\cite{xia2017deliberation, hu2020deliberation, hu2021transformer, hu2023scaling}.
Moreover, some works have employed pre-trained LMs, calculating the log-likelihood of each hypothesis with unidirectional models like GPT-2~\cite{radford2019language} or using pseudo-log-likelihood via bidirectional models such as BERT~\cite{devlin2018bert}.
LMs are also extensively employed for the error correction task across various languages, relying on the 1-best or N-best hypotheses generated by ASR models~\cite{leng2021fastcorrect,mani2020asr, leng2023softcorrect, ma2023n}.

Powered by LLMs, GER~\cite{chen2024hyporadise} integrates LLMs with ASR by providing hypotheses generated by ASR  as input to the LLM, using LoRA for fine-tuning, and adapting the LLM for the error correction task.
GER posits that the N-best hypotheses generated by ASR contain valuable information. Each hypothesis represents an independent textual representation of the input speech, potentially containing correct tokens for accurately predicting transcriptions.
GER begins by generating N-best hypotheses for the input speech through speech foundation models such as WavLM~\cite{chen2022wavlm} and Whisper~\cite{radford2023robust}. These hypotheses are subsequently processed by the LLM, which is fine-tuned using LoRA to establish the mapping between N-best hypotheses and the ground truth transcriptions.
Given the transcription $\mathbf{Y}$ corresponding to the speech and the N-best hypotheses $\mathcal{Y} = \left \{ \mathcal{Y}_\mathrm{1}, \mathcal{Y}_\mathrm{2}, \cdot \cdot \cdot ,\mathcal{Y}_\mathrm{N} \right \}$ generated by the speech foundation model, GER learns a hypotheses-to-transcription mapping in an auto-regressive manner:
\begin{equation}
    P(\hat{Y}_{l}|\hat{\mathbf{Y}}_{1:l-1},\mathcal{Y},\delta) = \text{LLM}_\delta(\hat{\mathbf{Y}}_{1:l-1}, \mathcal{Y}),
\end{equation}
where $\hat{Y}_{l}$ is the token to be predicted, $\hat{\mathbf{Y}}_{1:l-1}$ represents the preceding token sequence, and $\delta$ denotes the trainable parameters of LLM.
The GER model is optimized to maximize the estimated probability of the ground truth transcription with the standard cross-entropy (CE) loss:
\begin{gather}
    P(\mathbf{Y}|\mathcal{Y},\delta) = \prod_{l=1}^{L}P(Y_{l}|\mathbf{Y}_{1:l-1}, \mathcal{Y},\delta),\\
    \mathcal{L}_{\text{GER}} = \text{CrossEntropy}(P(\mathbf{Y}|\mathcal{Y},\delta), \mathbf{Y}),
\end{gather}
where $L$ means the length of the ground truth transcription.
\subsection{Multi-granularity Information Utilization}
The application of multi-granularity information for enhancing accented speech recognition performance has become commonplace.
Chan \textit{et al.}~\cite{chan2016online} integrates syllable and character multi-granularity information into an early attention-based ASR model with two decoders. Only the character decoder is employed for inference.
Other studies explore cascade speech-to-phoneme (S2P) and phoneme-to-word (P2W) schemes~\cite{zhou2018comparison, chen2018modular, zhou2018syllable, yuan2021decoupling, wang2021cascade}. Their experiments demonstrate that only text phoneme inputs for P2W lack adequate acoustic information compared to the speech embeddings generated by the encoder. Consequently, the two-stage independent structure accumulates errors, necessitating extensive P2W data to address the issue.
Zhang \textit{et al.}~\cite{zhang2021decoupling} proposes a method that merges a P2W model and an attention decoder during auto-regressive decoding, maximizing text phonemes and speech embedding for enhanced performance.
Yang \textit{et al.}~\cite{yang2022multi} leverages an attention decoder for P2W transcription. However, this concise and streamlined method is restricted to handling multi-granularity information of equal sequence lengths, exemplified by Chinese syllables and characters.
In contrast to approaches that directly apply fine-grained information to ASR, DIMNet~\cite{shao2023decoupling} incorporates it into the accent recognition (AR) task within a multi-task ASR-AR framework. The interaction between ASR and AR leads to significant performance gains in both tasks. Furthermore, it avoids the cascading S2P and P2W structures, thereby reducing error accumulation.

\subsection{LoRA}
LoRA~\cite{hu2021lora} represents an exemplary method for parameter-efficient fine-tuning, enabling the adaptation of large models to specific domains.
It achieves parameter efficiency by updating low-rank decomposition matrix pairs, ensuring the original weights remain fixed.
Specifically, for a given input vector $\mathbf{x}$ and a given linear layer with the weight matrix $\mathbf{W}_{\mathrm{0}}$, LoRA employs two low-rank decomposition matrix pairs, $\mathbf{A}$ and $\mathbf{B}$, with rank $r$, where $\mathbf{W}_{\mathrm{0}} \in \mathbb{R}^{d_{out} \times d_{in}}$, $\mathbf{A} \in \mathbb{R}^{r \times d_{in}}$, $\mathbf{B} \in \mathbb{R}^{d_{out} \times r}$, and $r \ll \min (d_{in}, d_{out})$.
Applying LoRA transforms the forward process of the given linear layer $\mathbf{l}=\mathbf{W}_{\mathrm{0}}\mathbf{x} + \mathbf{b}$ into the following formulation:
\begin{equation}
    \mathbf{l}=\mathbf{W}_{\mathrm{0}}\mathbf{x}+\Delta \mathbf{W}\mathbf{x} + \mathbf{b} = \mathbf{W}_{\mathrm{0}}\mathbf{x}+\frac{\alpha}{r}\mathbf{BAx} + \mathbf{b},\label{eq:lora}
\end{equation}
where the original weight $\mathbf{W}_{\mathrm{0}}$ and bias $\mathbf{b}$ remain fixed while low-rank matrices $\mathbf{A}$ and $\mathbf{B}$ are trainable during trainin.
Generally, the matrix $\mathbf{A}$ is initialized with a random Gaussian distribution, and matrix $\mathbf{B}$ starts from zero.
Scaling factor $\alpha/r$ controls the magnitude of adjustments to the original weights imposed by LoRA, whereas $\alpha$ and $r$ denote constants.

\subsection{MoLE}
MoE~\cite{shazeer2017outrageously, fedus2022review, zoph2022st, yuan2023moec, song2024u2++} framework expands model complexity and capacity by including diverse sub-network experts, each potentially addressing specific domain challenges.
An MoE layer utilizes a gating network router to manage $N$ independent experts $\left \{ \mathbf{E}^{i}  \right \}_{i=1}^{N}$. It generates a probability distribution through a trainable matrix and normalizes it via a softmax function to weigh the outputs of these experts.
For the given input vector $\mathbf{x}$, the output probability distribution $\mathbf{p}$ of the trainable router can be denoted as:
\begin{equation}
    \mathbf{p}=\text{Softmax}(\mathbf{W}_{g}\mathbf{x}),
\end{equation}
where $\mathbf{W}_{g}$ represents the trainable weights of the gating network router.
MoLE~\cite{liu2023moelora, feng2024mixture, zhu2023sira, li2024mixlora, yang2024moral, dou2024loramoe, mu2024hdmole, wu2024mixture} substitutes conventional dense layer experts with LoRA experts, rendering it a parameter-efficient fine-tuning approach.
The final output $\mathbf{m}$ from the MoLE layer is a weighted sum of the outputs from the top $K$ LoRA experts:
\begin{equation}
    \mathbf{m}=\sum_{i=1}^{K}\frac{\text{TopK}(\mathbf{p}^{i})}{ {\textstyle \sum_{j=1}^{K}} \text{TopK}(\mathbf{p}^{j})}\cdot \mathbf{E}^{i}(\mathbf{x}),  
\end{equation}
where $\mathbf{p}^{i}$ and $\mathbf{p}^{j}$ are the weights of $i\text{-th}$ and $j\text{-th}$ LoRA expert in MoLE layer, respectively.
The TopK(·) function selects the top $K$ with the highest weights, setting all other weights to zero.
The weights selected by the TopK(·) function are normalized to ensure their sum equals one.
The computation process of each LoRA expert is described in Eq.~\ref{eq:lora}.

Various studies focus on exploring multiple aspects of MoLE, including the integration between LoRA experts and the original weight matrix (e.g., plugin or fusion), the number of experts selected (e.g., Top-1, Top-K, or Top-All), the granularity of expert selection (e.g., token-level or task-level), and the implementation of experts load balancing loss.
MOELoRA~\cite{liu2023moelora} employs LoRA experts as plugins for the LLM, selecting the Top-1 expert based on task IDs to perform medical multi-task learning.
LoRAMoE~\cite{dou2024loramoe} embeds LoRA experts as plugins in the feed-forward networks (FFN) of the LLM, selecting Top-All experts based on input tokens to prevent world knowledge forgetting within the LLM.
MixLoRA~\cite{li2024mixlora} fuses LoRA experts with the LLM, selecting Top-K experts based on input tokens. It employs auxiliary loss to mitigate expert load imbalance, achieving a parameter-efficient MoLE.
Despite advancements, MoLE still faces unresolved issues regarding expert load imbalance and the rigidity of expert selection.

\section{Proposed Methods}
This section outlines the three-stage training strategy for obtaining mono-accent LoRA experts from multi-modal GER. Following this, we describe the HDMoLE method for combining these mono-accent LoRA experts and explain the underlying mechanism of multi-granularity GER.

\subsection{Mono-accent LoRA Experts from Multi-modal GER}
The vanilla GER relies exclusively on N-best word-level hypotheses to predict transcriptions, neglecting the pronunciation information in accented speech.
Multi-modal GER improves vanilla GER by integrating pronunciation information from accented speech. It empowers the LLM to leverage pronunciation information from the speech modality and semantic information from the text modality to achieve more precise transcription predictions for accented speech.
\begin{figure}[t]
  \centering
  \includegraphics[width=1.15\linewidth]{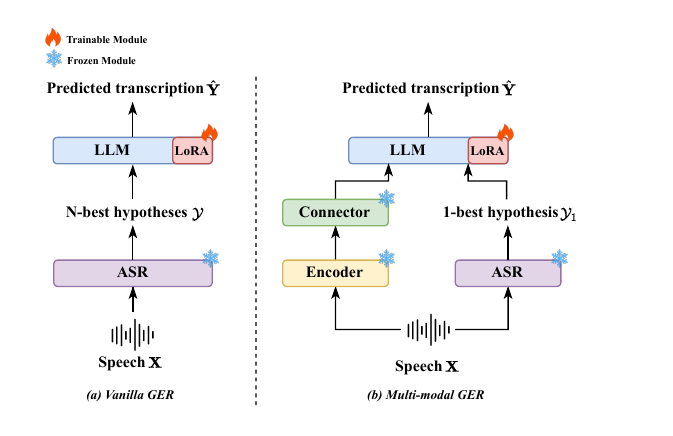}
  \caption{Overview of vanilla GER and our proposed multi-modal GER. In this figure, ``Encoder" means the speech foundation encoder, and ``ASR" means the speech foundation model. \textit{\textbf{Left}}: vanilla GER employs N-best word-level hypotheses generated by speech foundation models to predict ground-truth transcription; \textit{\textbf{Right}}: multi-modal GER integrates pronunciation information from speech embeddings and semantic information from 1-best word-level hypotheses to predict ground-truth accented speech transcription.}
  \label{fig:fig1}
\end{figure}
Fig.~\ref{fig:fig1} provides an overview of the vanilla GER and multi-modal GER.
The multi-modal GER utilizes the vanilla Whisper-large-v3 to produce 1-best word-level hypotheses and employs its encoder as the speech foundation encoder.
Convolution and linear projection layers serve as the connector module, with LoRA fine-tuning the LLM to map the speech embedding and hypothesis pair to the corresponding transcription.
Due to computational resource constraints, 1-best word-level hypotheses are adopted to minimize input sequence length, delivering highly commendable results nonetheless.
Additionally, the vanilla Whisper encoder can produce high-quality speech embeddings.
Within the connector module, convolution layers perform downsampling to reduce the length of speech embeddings, while linear projection layers ensure alignment between the speech and text modalities.
\begin{figure*}[!t]
  \centering
  \includegraphics[width=1.0\linewidth]{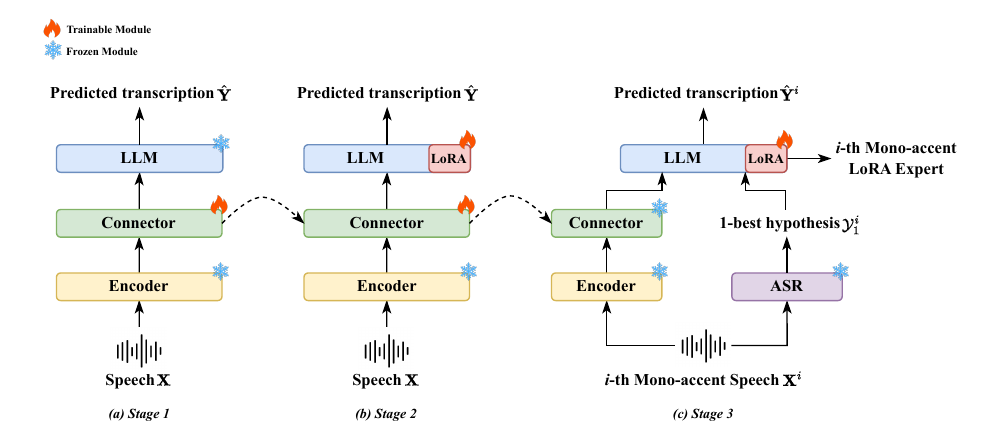}
  \caption{Overview of the three-stage training strategy for obtaining mono-accent LoRA experts from multi-modal GER. In this figure, ``Encoder" means the vanilla Whisper-large-v3 encoder, ``ASR" means the vanilla Whisper-large-v3 model, and the dashed arrows indicate the source of model parameter initialization. \textbf{\textit{Stage 1}}: freeze the speech foundation encoder and the LLM, training only the connector module; \textbf{\textit{Stage 2}}: load the connector module from the last stage, training both connector module and LoRA part of the LLM; \textbf{\textit{Stage 3}}: load the connector module from the previous stage, training only the LoRA part of the LLM using $i$-th mono-accent speech data and corresponding 1-best hypotheses from vanilla Whisper-large-v3 model greedy search decoding.}
  \label{fig:fig2}
\end{figure*}
As illustrated in Fig.~\ref{fig:fig2}, we conduct a three-stage training strategy to obtain mono-accent LoRA experts from multi-modal GER.
In the first stage, the speech foundation encoder and LLM are frozen, and only the connector module is trained using all accented speech data.
In the second stage, the parameters of the connector module from the first stage are loaded, and both the connector module and the LoRA part of the LLM are trained using all accented speech data.
Appropriately initialization is essential as the connector module remains fixed during the third stage of training multi-modal GER.
During the first two training stages, the connector module acquires the ability to align the speech and text modalities while also adapting the LoRA portion of the LLM.
Consequently, we initialize the connector module of the multi-modal GER using the parameters of the connector module obtained from the first two training stages.
In the third stage, we train the LoRA components of the multi-modal GER using mono-accent speech data, thereby acquiring mono-accent LoRA experts.
Assuming there are $N$ accents, for the $i$-th accent, the input speech is $\mathbf{X}^{i}$, the corresponding 1-best hypothesis is $\mathcal{Y}_{1}^{i}$, and the transcription is $\mathbf{Y}^{i}$.
In multi-modal GER, the speech foundation encoder first transforms the input accented speech into speech embeddings $\mathbf{H}_\mathrm{enc}^{i}$, denoted as:
\begin{equation}
    \mathbf{H}_\mathrm{enc}^{i} = \text{Encoder}(\mathbf{X}^{i}).
\end{equation}
Then, $\mathbf{H}_\mathrm{enc}^{i}$ are aligned to the text modality via the frozen connector module, producing feature sequences $\mathbf{H}_\mathrm{con}^{i}$ with dimensions matching those required by the LLM input, which can be written as:
\begin{equation}
    \mathbf{H}_\mathrm{con}^{i} = \text{Connector}(\mathbf{H}_\mathrm{enc}^{i}).
\end{equation}
Meanwhile, the 1-best hypothesis is tokenized by the LLM tokenizer:
\begin{equation}
    \mathbf{H}_\mathrm{hyp}^{i} = \text{Tokenizer}(\mathcal{Y}_{1}^{i}).
\end{equation}
Next, we concatenate the modality-aligned speech embeddings, 1-best hypothesis embeddings, and prompt embeddings along the length dimension to construct the input for the LLM $\mathbf{H}^{i}$, which can be represented as:
\begin{equation}
    \mathbf{H}^{i} = \text{Concat}(\mathbf{H}_\mathrm{hyp}^{i},\mathbf{H}_\mathrm{pro}^{i},\mathbf{H}_\mathrm{enc}^{i}),
\end{equation}
where $\mathbf{H}_\mathrm{pro}^{i}$ is the trainable prompt embeddings.
After that, LLM predicts the probability distribution for the next token auto-regressively:
\begin{equation}
    P(\hat{Y}^{i}_{l}|\hat{\mathbf{Y}}^{i}_{1:l-1},\mathbf{H}^{i},\delta^{i}) = \text{LLM}_{\delta^{i}}(\hat{\mathbf{Y}}^{i}_{1:l-1}, \mathbf{H}^{i}),
\end{equation}
where $\hat{Y}_{l}^{i}$ is the token to be predicted, $\hat{\mathbf{Y}}_{1:l-1}^{i}$ represents the predicted token sequence and $\delta^{i}$ denotes the parameters of LoRA experts for the $i$-th accent.
The training objective of the multi-modal GER model is to maximize the estimated probability of the ground truth transcription with the CE loss:
\begin{gather}
     P(\mathbf{Y}^{i}|\mathbf{H}^{i},\delta^{i}) = \prod_{l=1}^{L}P(Y_{l}^{i}|\mathbf{Y}_{1:l-1}^{i}, \mathbf{H}^{i},\delta^{i}),\\
    \mathcal{L}^{i}_{\text{Multi-modal}} = \text{CrossEntropy}(P(\mathbf{Y}^{i}|\mathbf{H}^{i},\delta^{i}), \mathbf{Y}^{i}),
\end{gather}
where $L$ means the length of the ground truth transcription.
Ultimately, we obtain mono-accent LoRA experts from multi-modal GER through the previously described three-stage training strategy.	
\begin{figure*}[!t]
  \centering
  \includegraphics[width=1.0\linewidth]{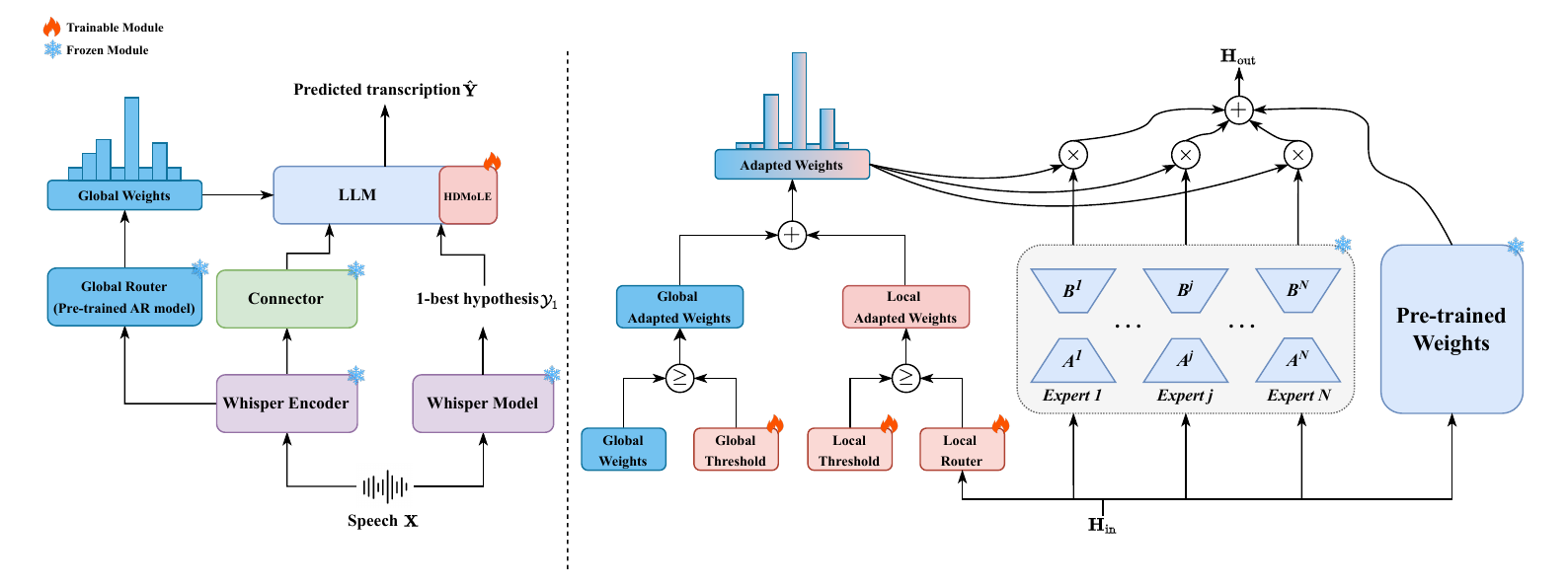}
  \caption{Overview of the proposed HDMoLE for LoRA experts combination. \textbf{\textit{Left}}: HDMoLE is integrated into the multi-modal GER model, where a pre-trained accent recognition (AR) model provides global weights for HDMoLE; \textbf{\textit{Right}}: each HDMoLE layer employs hierarchical routing and dynamic thresholds to obtain adapted weights for combining mono-accent LoRA experts.}
  \label{fig:fig3}
\end{figure*}
\subsection{Hierarchical Routing in HDMoLE}
The original purpose of MoE is for routers to allocate weights to each expert based on the input sample, prioritizing experts specialized in the input domain and thereby creating a clear correspondence between experts and domains.
In practice, routers in MoE layers frequently converge to a state where they consistently assign large weights to high-performing experts from earlier stages, resulting in only a limited number of experts exerting a significant influence.
The experts prioritized by the MoE router with higher weights tend to remain unchanged, regardless of variations in the input domains.
As a result, the disproportionate reliance on several experts diminishes the clarity of the correspondence between experts and domains.
Some works~\cite{shazeer2017outrageously, wu2024mixture, li2024mixlora} employ expert utilization balance loss to promote a more balanced weight allocation across the experts by the routers.
However, the loss only ensures balanced expert utilization by reducing weight disparities without directly establishing an explicit mapping between experts and their corresponding domains.
To mitigate this limitation, we propose a hierarchical routing strategy in HDMoLE that establishes a clear correspondence between LoRA experts and domains.

Fig.~\ref{fig:fig3} shows the overview and details of our proposed HDMoLE.
Specifically, hierarchical routing consists of global and local routing with a pre-trained AR model as the global router and individual HDMoLE layer routers as local routers.
The global routing explicitly directs each LoRA expert to specialize in a particular accent domain, thereby clarifying the correspondence between LoRA experts and accent domains and designating them as domain-specific LoRA experts.
At the same time, the local routing implicitly learns the associations among LoRA experts. It guides the LoRA experts within the HDMoLE layer to collaborate across various accent domains through a learnable gating network router.
The input speech $\mathbf{X}$ generates global weights $\mathbf{P}_\mathrm{g}$ through the Whisper speech foundation encoder and the global router, denoted as:
\begin{gather}
    \mathbf{H}_\mathrm{enc} = \text{Encoder}(\mathbf{X}), \\
    \mathbf{P}_\mathrm{g} = \text{Softmax}(\text{Router}_\text{Global}(\mathbf{H}_\mathrm{enc})),
\end{gather}
where the frozen global router combines GRU~\cite{cho2014properties} and linear projection layers.
Then, the global weights $\mathbf{P}_\mathrm{g}$ are input into each HDMoLE layer.
Moreover, the input hidden states $\mathbf{H}_\mathrm{in}$ produce local weights $\mathbf{P}_\mathrm{l}$ through the local router in each HDMoLE layer, which can be written as:
\begin{equation}
    \mathbf{P}_\mathrm{l} = \text{Softmax}(\text{Router}_\text{Local}(\mathbf{H}_\mathrm{in})),
\end{equation}
where the local router is a trainable linear projection layer.
\subsection{Dynamic Thresholds in HDMoLE}
Standard MoE layers commonly employ static expert selection strategies, such as the Top-K expert selection, in which the router chooses the $K$ experts with the highest weights.
As each MoE layer concentrates on different domain aspects, the number of experts chosen varies across MoE layers.

Therefore, we propose a dynamic expert selection strategy with trainable thresholds for each HDMoLE layer, offering a more flexible alternative to the static expert selection strategies.
With the dynamic thresholds strategy, each HDMoLE layer can adaptively select the LoRA experts to be activated based on the requirements.
This strategy follows a rule whereby a LoRA expert is selected if its weight exceeds the dynamic thresholds.
Proper initialization of the dynamic threshold is crucial, as an excessively high threshold may cause all LoRA expert weights to fall below it, leaving no LoRA experts selected.
Each HDMoLE layer's dynamic thresholds are initialized at $1/N$ to guarantee that at least one LoRA expert is selected. Two independent dynamic thresholds are assigned to the global and local weights.
We can obtain the global adapted weights $\mathbf{P}_\mathrm{ga}$ through the global threshold $\tau_{g}$, which can be impressed as follows:
\begin{equation}
    \mathbf{P}_\mathrm{ga} = \frac{\mathbb{E}(\mathbf{P}_{\mathrm{g} }\ge \tau_{g} )\cdot \mathbf{P}_{\mathrm{g} }}{ {\textstyle \sum_{j=1}^{N}} \mathbb{E}(\mathbf{P}_{\mathrm{g} }^{j}\ge \tau_{g} )\cdot \mathbf{P}_{\mathrm{g} }^{j}} \cdot \tau_{g},
\end{equation}
where $\mathbb{E}$(condition) is assigned a value of one if it is true and zero if it is false. $\mathbf{P}_{\mathrm{g}}^{j}$ represents the global weight of the $j$-th LoRA expert.
Additionally, scaling the global adapted weights by $\tau_{g}$ guarantees that the $\tau_{g}$ remains trainable during backpropagation.
In the same way, the local threshold $\tau_{l}$ allows us to obtain the local adapted weights $\mathbf{P}_\mathrm{la}$, which can be impressed as follows:
\begin{equation}
    \mathbf{P}_\mathrm{la} = \frac{\mathbb{E}(\mathbf{P}_{\mathrm{l} }\ge \tau_{l} )\cdot \mathbf{P}_{\mathrm{l} }}{ {\textstyle \sum_{j=1}^{N}} \mathbb{E}(\mathbf{P}_{\mathrm{l} }^{j}\ge \tau_{l} )\cdot \mathbf{P}_{\mathrm{l} }^{j}} \cdot \tau_{l},
\end{equation}
where $\mathbf{P}_{\mathrm{l}}^{j}$ represents the local weight of the $j$-th LoRA expert.
The final adapted weights $\mathbf{P}_\mathrm{a}$ are the sum of the global and local adapted weights, which can be defined as follows:
\begin{equation}
    \mathbf{P}_\mathrm{a} = \mathbf{P}_\mathrm{ga} + \mathbf{P}_\mathrm{la}.
\end{equation}
\subsection{Mixture of LoRA Experts in HDMoLE}
In MoLE, the combination of LoRA experts can be primarily classified into two categories. 
One involves simultaneously training all LoRA experts and the local router from scratch~\cite{liu2023moelora, li2024mixlora, yang2024moral, dou2024loramoe}.  
The other loads pre-trained LoRA expert parameters, keeping the LoRA experts frozen all the time and training only the local router from scratch~\cite{wu2024mixture}, further reducing the number of trainable parameters while preserving the capabilities of the LoRA experts.
Notably, since accents reflect variations in pronunciation within the same language, a carefully crafted combination of pre-trained mono-accent LoRA experts across accents can lead to superior outcomes in accented speech recognition.
HDMoLE combines mono-accent LoRA experts using adapted weights derived from hierarchical routing and dynamic thresholds.
Specifically, in each HDMoLE layer, hierarchical routing and dynamic thresholds determine the adapted weights $\mathbf{P}_\mathrm{a}$ based on the input hidden states $\mathbf{H}_\mathrm{in}$. 
These adapted weights are then applied to combine the outputs of the various LoRA experts, formulated as:
\begin{equation}
    \mathbf{H}_\mathrm{out} = \mathbf{W}_{0}\mathbf{H}_\mathrm{in} + \frac{\alpha}{r} \sum_{j=1}^{N} \mathbf{P}_{\mathrm{a}}^{j} \cdot \mathbf{B}^{j}\mathbf{A}^{j}\mathbf{H}_{\mathrm{in}} + \mathbf{b},
\end{equation}
where low-rank matrices $\mathbf{A}^{j}$ and $\mathbf{B}^{j}$ are loaded from the pre-trained $j$-th mono-accent LoRA expert $\delta^{j}$, while the original LLM layer with the weight matrix $\mathbf{W}_{0}$ and bias $\mathbf{b}$.
Each LoRA expert's output is scaled by its corresponding adapted weight to achieve a weighted summation among the LoRA experts.
When a specific LoRA expert's global and local weights are simultaneously lower than the global and local thresholds, the LoRA expert's weight in the final adapted weights $\mathbf{P}_\mathrm{a}$ becomes zero.
This weighted summation is subsequently scaled by $\alpha/r$, maintaining consistency with the scaling applied during the acquisition of mono-accent LoRA experts, where LoRA rank $r$ and LoRA alpha $\alpha$ are pre-defined constants.
Finally, HDMoLE optimizes the parameters of local routers and dynamic thresholds through the CE loss:
\begin{equation}
    \mathcal{L}_{\text{HDMoLE}} = \text{CrossEntropy}(P(\mathbf{Y}|\mathbf{H},\sigma), \mathbf{Y}),
\end{equation}
where $\mathbf{Y}$ is the ground truth transcriptions, $\mathbf{H}$ is the concatenation of speech embeddings, 1-best hypothesis embeddings, and prompt embeddings, and $\sigma$ is the parameters of local routers and thresholds.
\begin{figure}[!t]
  \centering
  \includegraphics[]{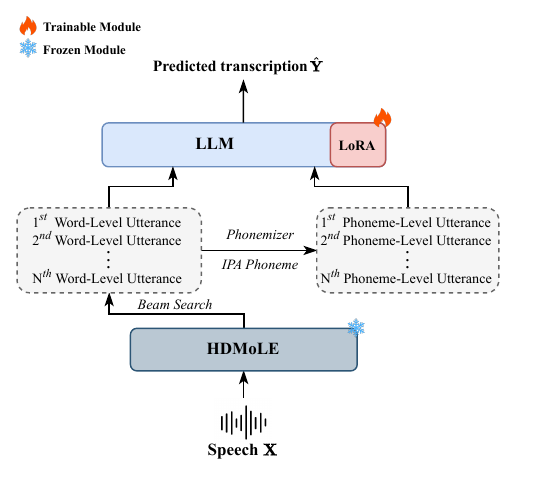}
  \caption{Overview of the multi-granularity GER. The N-best word-level and phoneme-level hypotheses generated by the HDMoLE model beam search decoding are inputs to the multi-granularity GER, enabling the LLM to produce the final predicted transcription.}
  \label{fig:fig4}
\end{figure}
\subsection{Multi-Granularity GER}
Multi-granularity GER improves vanilla GER by incorporating pronunciation information from N-best phoneme-level hypotheses. This allows the LLM to leverage the text modality pronunciation and semantic information to produce the final accented speech transcription.
Finer-grained linguistic information facilitates the activation of LLMs' more comprehensive text-processing capabilities.
Phonemes, as fine-grained linguistic units associated with pronunciation, provide the LLM with a deeper understanding of word pronunciation patterns, enhancing word-level error correction.
Conversely, as coarse-grained linguistic units related to semantics, words allow the LLM to better comprehend the semantic content of hypotheses, improving hypothesis-level error correction.
MMGER~\cite{mu2024mmger} performs fine-grained error correction by force-aligning speech embeddings and 1-best hypotheses, concatenating speech embeddings and their corresponding hypothesis character embeddings along the feature dimension.
Furthermore, since the 1-best hypothesis loses its semantic information during the forced alignment process, the original 1-best hypothesis is reintroduced to supplement the semantic information, enabling multi-granularity error correction.
In contrast to MMGER, we perform multi-granularity error correction through phoneme-level and word-level hypotheses.

Fig.~\ref{fig:fig4} provides an overview of multi-granularity GER.
We decode the trained HDMoLE model using beam search to obtain the N-best word-level hypotheses to implement multi-granularity GER. 
Subsequently, we convert the word-level hypotheses into phoneme-level hypotheses using the phonemizer toolkits\footnote{https://github.com/bootphon/phonemizer}, representing the phonemes in the International Phonetic Alphabet (IPA).
Finally, we concatenate the N-best word-level hypotheses and phoneme-level hypotheses and input them into the LLM.
Through LoRA fine-tuning, the LLM learns the mapping between multi-granularity hypotheses and the transcriptions.
For the input speech $\textbf{X}$, the corresponding N-best word-level hypotheses $\mathcal{Y}$ and phoneme-level hypotheses $\mathcal{P}$, the multi-granularity hypotheses are tokenized by the LLM tokenizer:
\begin{gather}
    \mathbf{H}_\mathrm{word} = \text{Tokenizer}(\mathcal{Y}), \\
    \mathbf{H}_\mathrm{phoneme} = \text{Tokenizer}(\mathcal{P}).
\end{gather}
Then, we concatenate the multi-granularity hypotheses embeddings and input them to the LLM. 
The next text token probability through LoRA fine-tuning can be impressed as:
\begin{gather}
    \mathbf{E} = \text{Concat}(\mathbf{H}_\mathrm{word},\mathbf{H}_\mathrm{phoneme}), \\
    P(\hat{Y}_{l}|\hat{\mathbf{Y}}_{1:l-1},\mathbf{E},\omega) = \text{LLM}_{\omega}(\hat{\mathbf{Y}}_{1:l-1}, \mathbf{E}),
\end{gather}
where $\hat{Y}_{l}$ is the token to be predicted, $\hat{\mathbf{Y}}_{1:l-1}$ represents the preceding token sequence, and $\omega$ is the trainable LoRA parameters of multi-granularity GER.
The loss of multi-granularity GER is CE loss:
\begin{gather}
    P(\mathbf{Y}|\mathbf{E},\omega) = \prod_{l=1}^{L}P(Y_{l}|\mathbf{Y}_{1:l-1}, \mathbf{E},\omega),\\
    \mathcal{L}_{\text{Multi-granularity}} = \text{CrossEntropy}(P(\mathbf{Y}|\mathbf{E},\omega), \mathbf{Y}),
\end{gather}s
where $\mathbf{Y}$ is the ground truth transcription of length $L$.

\section{Experimental Setups}
This section first provides a comprehensive description of the datasets. We then explain the setup details for the multi-modal GER, HDMoLE, and multi-granularity GER models and describe the configurations applied during the training and inference processes.
\begin{table}[h]
    \caption{The duration (Hours) of English with 9 accents in other datasets and our combined dataset. ``CV 19.0" means the Common Voice Corpus 19.0.}
    \label{tab:tabel1}
    \centering
\begin{tabular}{l|ccccccccc}
\toprule
\textbf{Accent} & \textbf{AU} & \textbf{CA}  & \textbf{CN} & \textbf{DE} & \textbf{IN}  & \textbf{JP} & \textbf{UK}  & \textbf{US}  & \textbf{ZA} \\ \midrule
\textbf{CV 19.0} & 76 & 109 & 0 & 98 & 161 & 0 & 215 & 609 & 38 \\
\textbf{VCTK} & 1 & 3 & 0 & 0 & 1 & 0 & 13 & 8 & 1 \\
\textbf{AESRC} & 0 & 2 & 22 & 0 & 22 & 22 & 23 & 22 & 0 \\ \midrule
\textbf{Ours}  & 77 & 114 & 22 & 98 & 184 & 22 & 251 & 639 & 39 \\ \bottomrule
\end{tabular}
\end{table}
\subsection{Datasets}
We combine three English datasets with accent annotations to evaluate our proposed methods: Common Voice Corpus 19.0~\cite{ardila2019common}, VCTK\footnote{https://datashare.ed.ac.uk/handle/10283/3443}, and AESRC~\cite{shi2021accented}.

The Common Voice Corpus is a comprehensive multilingual collection of transcribed read speech intended for speech technology research and development.
The English dataset in Common Voice Corpus 19.0 comprises 2,658 hours of validated data from 93,896 English speakers, from which we select subsets with accent annotations for our experiments.
The VCTK dataset includes approximately 39 hours of speech data provided by 110 English speakers with various accent annotations.
Each speaker reads about 400 sentences selected from a newspaper, the rainbow passage, and an elicitation paragraph for the speech accent archive.
The AESRC dataset originates from the Interspeech2020 Accented English Speech Recognition Challenge and comprises about 180 hours of speech data encompassing 10 types of English accents.
The speakers read sentences that cover common conversational topics and human-computer interaction commands.

For our combined multi-accent English dataset, we select 9 prominent and prevalently used English accents from Common Voice Corpus 19.0, VCTK, and AESRC: Australian (AU), Canadian (CA), Chinese (CN), German (DE), Indian (IN), Japanese (JP), British (UK), American (US), and South African (ZA).	
Table~\ref{tab:tabel1} shows the duration of our combined multi-accent English dataset with 9 accents.
\subsection{Setups}
For multi-modal GER and HDMoLE, the speech foundation encoder is derived from the vanilla Whisper-large-v3 encoder.
The connector module consists of two convolution layers, two linear projection layers, and ReLU activation functions.
Additionally, 1-best word-level hypotheses are obtained through greedy search decoding executed by the vanilla Whisper-large-v3 on all speech data in our combined dataset.
Moreover, we employ LLaMA-3.2-3B as the LLM, with each LoRA expert configured with a rank of 32, an alpha value of 8, and a dropout rate of 0.05.
The pre-trained AR model consists of a frozen vanilla Whisper-large-v3 encoder, a 4-layer GRU with a hidden size of 256 and a dropout rate of 0.1, a 3-layer linear projection with a hidden size of 4096, and ReLU activation functions.
Each HDMoLE layer in the LLM includes trainable parameters consisting of two dynamic thresholds initialized at $1/9$ and a linear projection layer as the local router that maps the input feature dimensions to the number of accents.

For multi-granularity GER, the N-best word-level hypotheses are generated by conducting beam search decoding on all speech data in our combined dataset using the trained HDMoLE model.
The N-best word-level hypotheses are converted into corresponding IPA phoneme-level hypotheses using the phonemizer toolkits.
We employ LLaMA-3.1-8B as the LLM, with the LoRA configuration set to a rank of 64, an alpha value of 16, and a dropout rate of 0.05.
In all experiments involving multi-modal GER, HDMoLE, and multi-granularity GER, the prompt embeddings are trained from scratch and have a length of 50.
\begin{table*}[t]
    \caption{The WER(\%) results of our proposed methods on our combined multi-accent English test sets. ``\#TrP (M)" refers to the number of trainable parameters. ``\#TotP (M)" denotes the number of total parameters. ``FLOPs (G)" represents the inference cost of one text token or speech frame (10 ms). ``AVG" means the average WER across the test sets of 9 accents.}
    \label{tab:tabel2}
    \centering
\scalebox{0.95}{
\begin{tabular}{lccccccccccccc}
\toprule
\multirow{2}{*}{\textbf{Model}} & \multirow{2}{*}{\textbf{\#TrP (M)}} & \multirow{2}{*}{\textbf{\#TotP (M)}} & \multirow{2}{*}{\textbf{FLOPs (G)}} & \multicolumn{10}{c}{\textbf{Test WER (\%)}}                                                                                                                   \\ \cmidrule{5-14} 
                                &                                     &                                      &                            & \textbf{AVG}  & \textbf{AU}   & \textbf{CA}   & \textbf{CN}   & \textbf{DE}   & \textbf{IN}   & \textbf{JP}   & \textbf{UK}   & \textbf{US}   & \textbf{ZA}   \\ \midrule
Vanilla Whisper~\cite{radford2023robust}                 & -                                    & 1541.62                              & 8.29                       & 6.34          & 6.08          & 3.37          & 6.04          & 5.56          & 7.19          & 14.99         & 2.02          & 6.84          & 13.79         \\
Qwen2-Audio~\cite{chu2024qwen2}                     & -                                    & 8397.09                              & 21.85                      & 5.81          & 5.35          & 3.82          & 5.22          & 3.89          & 7.88          & 10.89         & 2.21          & 6.12          & 10.54         \\
FireRedASR-LLM-L~\cite{xu2025fireredasr}                & -                                    & 8509.09                              & 22.37                      & 7.00          & 6.74          & 4.13          & 5.11          & 4.81          & 11.11         & 13.13         & 2.62          & 7.36          & 17.05         \\ \midrule
Fully fine-tuned Whisper        & 1541.62                             & 1541.62                              & 8.29                       & 3.29          & 3.31          & 1.93          & 4.46          & 3.72          & 4.77          & 3.40          & 1.43          & 3.60          & 11.28         \\
Vanilla GER (LLaMA-3.2-3B)      & 97.44                               & 3310.01                              & 6.62                       & 3.26          & 3.09          & 1.93          & 4.46          & 3.70          & 4.72          & 3.40          & 1.41          & 3.56          & 11.30         \\
Vanilla GER (LLaMA-3.1-8B)      & 167.95                              & 8198.03                              & 15.35                      & 3.22          & 2.98          & 1.95          & 4.43          & 2.96          & 4.68          & 3.41          & 1.36          & 3.46          & 10.92         \\ \midrule
Multi-modal GER                 & 83.86                               & 3966.95                              & 13.02                      & 3.20          & 3.31          & 1.90          & 4.34          & 3.70          & 4.56          & 3.38          & 1.38          & 3.52          & 11.49         \\
\hspace{1em}+ Multi-granularity GER         & 167.95                              & 8198.03                              & 15.35                      & 2.62          & 2.54          & 1.40          & 4.05          & 1.85          & 3.57          & 3.36          & 0.95          & 2.70          & 10.92         \\
Mono-accent Multi-modal GER     & 439.29                              & 4322.38                              & 13.02                      & 2.92          & 2.98          & 1.69          & 4.39          & 2.22          & 3.98          & 3.38          & 1.18          & 3.08          & 11.11         \\
\hspace{1em}+ Multi-granularity GER         & 167.95                              & 8198.03                              & 15.35                      & 2.43          & 2.21          & 1.32          & 3.88          & 1.48          & 3.24          & 3.29          & 0.85          & 2.46          & 10.83         \\
HDMoLE                          & 6.72                                & 4329.10                              & 13.70                      & 2.49          & 2.21          & 1.33          & 3.90          & 1.48          & 3.35          & 3.34          & 0.80          & 2.60          & 9.00          \\
\hspace{1em}+ Multi-granularity GER         & 167.95                              & 8198.03                              & 15.35                      & \textbf{2.07} & \textbf{1.22} & \textbf{0.92} & \textbf{3.44} & \textbf{1.11} & \textbf{2.91} & \textbf{2.71} & \textbf{0.54} & \textbf{2.23} & \textbf{8.62} \\ \bottomrule
\end{tabular}}
\end{table*}

\subsection{Training and Inference Configurations}
All multi-modal GER and HDMoLE experiments include speed perturbation and SpecAugment for data augmentation, while the multi-granularity GER experiments do not include data augmentation.
Each model is trained for 6 epochs, and the best 3 checkpoints are averaged for the final inference.
We use 8 40GB A100 GPUs for all experiments.
A warm-up scheduler adjusts the learning rate, peaking at $1e-4$ with 2500 steps.
In all inference processes of our experiments, the LLM generates transcriptions without employing sampling methods (i.e., do\_sample=False), with the temperature, repetition penalty, and length penalty all set to 1, and we use the greedy search decoding strategy for the inference.
Seven modules in each LLM layer, including q\_proj, k\_proj, v\_proj, o\_proj, up\_proj, gate\_proj, and down\_proj, are subject to the LoRA and various MoLE methods.
\section{Experimental Results and Analysis}
This section compares our proposed methods with existing methods, accompanied by ablation studies, to examine their efficacy. Furthermore, we provide a comprehensive analysis of the experimental results.
\subsection{Results of Our Proposed Methods}
Table~\ref{tab:tabel2} presents the comparison WER results of vanilla Whisper-large-v3, Qwen2-Audio~\cite{chu2024qwen2}, FireRedASR-LLM-L~\cite{xu2025fireredasr}, fully fine-tuned Whisper-large-v3, vanilla GER, and our proposed methods.
Qwen2-Audio\footnote{https://github.com/QwenLM/Qwen2-Audio} is a large-scale audio-language model capable of processing various audio signals and performing audio analysis.
Furthermore, FireRedASR\footnote{https://github.com/FireRedTeam/FireRedASR} is a family of open-source, industrial-grade ASR models that support Mandarin, Chinese dialects, and English, achieving new state-of-the-art results on public ASR benchmarks.
FireRedASR-LLM-L adopts an Encoder-Adapter-LLM framework, harnessing the power of LLM to focus specifically on the ASR task.
Both Qwen2-Audio and FireRedASR-LLM-L perform high recognition accuracy in general speech conditions.
We obtain the inference results by following the official evaluation procedures.
Although these two models are trained on large-scale data, they still perform moderately in multi-accent scenarios.
Moreover, the large-scale ASR foundation model, vanilla Whisper, encounter notable challenges under accented speech conditions; however, fully fine-tuned Whisper mitigates these challenges to some extent.
Whether LoRA rank 64 is applied to LLaMA-3.1-8B or LLaMA-3.2-3B, the vanilla GER demonstrates performance comparable to fully fine-tuned Whisper because we reproduce GER by fine-tuning the LLM with LoRA to learn the mapping between the N-best hypotheses generated by the fully fine-tuned Whisper and the ground truth transcriptions.
In contrast, the multi-modal GER using our entire combined dataset achieves substantial improvements over the vanilla Whisper and GER, owing to the incorporation of the 1-best hypotheses and speech embeddings. This effectively activated the LLM's error correction potential to predict accurate transcriptions for accented speech.
The mono-accent multi-modal GER surpasses multi-modal GER because it optimizes distinct LoRA parameters customized for specific accent conditions, producing mono-accent LoRA experts.
Combining these mono-accent LoRA experts through HDMoLE achieves enhanced improvements, taking advantage of the inherent complementarity of diverse accents within the same language.
To validate the effectiveness of HDMoLE, we report the WER results using only multi-modal GER and multi-granularity GER.
Finally, employing multi-granularity GER to correct the word-level and phoneme-level decoding results of the trained HDMoLE model further achieves the best performance.
Although our methods involve substantially fewer trainable parameters than fully fine-tuned Whisper, they significantly improve performance.
In addition, we also report the computational cost (FLOPs) for each method during inference, which corresponds to the processing of one speech frame (10 ms) or one text token.
Since multi-modal GER requires the speech encoder to provide speech embeddings, it incurs a higher computational cost compared to the vanilla GER (13.02 G vs. 6.62 G).
Although HDMoLE combines multiple mono-accent LoRA experts, the inference computational cost does not increase significantly (13.02 G vs. 13.70 G).
Multi-granularity GER only adds phoneme-level hypotheses during training, so its inference computational cost is the same as that of the vanilla GER (15.35 G).
\begin{table}[h]
    \caption{The WER(\%) results of our proposed HDMoLE method and other MoLE methods on our combined multi-accent English test sets. ``\#TrP (M)" refers to the number of trainable parameters, ``Top-K" denotes the top K experts to be selected, ``Local Weights Level" represents utterance or frame level local weights, and ``Test WER" means the average WER across the test sets of 9 accents.}
    \label{tab:tabel3}
    \centering
\scalebox{0.85}{
\begin{tabular}{lcccc}
\toprule
\textbf{Model} & \textbf{\#TrP (M)} & \textbf{Top-K} & \textbf{Local Weights Level} & \textbf{Test WER (\%)} \\ \midrule
MOELoRA~\cite{liu2023moelora}        & 0.84           & 1              & Utterance-level             & 3.55         \\
MoRAL~\cite{yang2024moral}          & 1.48           & 2              & Frame-level                 & 3.47         \\
MoLE~\cite{wu2024mixture}           & 2.21           & 3              & Utterance-level             & 3.40         \\
LoRAMoE~\cite{dou2024loramoe}        & 6.72           & All            & Frame-level                 & 3.22         \\
MoA~\cite{feng2024mixture}            & 6.72           & All            & Utterance-level             & 3.33         \\ \midrule
HDMoLE         & 6.72           & -              & Utterance-level             & 2.57         \\
HDMoLE         & 6.72           & -              & Frame-level                 & \textbf{2.49}         \\ \bottomrule
\end{tabular}}
\end{table}
\subsection{Results of HDMoLE and Other MoLE Methods}
Table~\ref{tab:tabel3} illustrates the WER results of our proposed HDMoLE method and other MoLE methods.
Although compared MoLE methods are not open-sourced and applied initially in various domains, the primary distinctions among them are the static Top-K expert selection, with varying values for $K$, and the level at which the local router assigns weights, whether at the utterance or frame level.
By pooling the input hidden states along the length dimension before inputting them into the local router, utterance-level local weights can be obtained.
We combine the mono-accent LoRA experts strictly according to the settings used in the compared MoLE methods to reproduce the results.
The HDMoLE method consistently demonstrates superior performance compared to other MoLE methods.
Under the same number of trainable parameters, HDMoLE achieves better WER results than other MoLE methods.
Meanwhile, we present the impact of different local weight levels on HDMoLE, and it is evident that frame-level local weights are more suitable for HDMoLE.
\begin{table}[h]
    \caption{Ablation study WER(\%) results for multi-modal GER with different granularities of hypotheses on our combined multi-accent English test sets. ``Speech Emb." refers to speech embeddings, ``Hyp." represents the granularity of 1-best hypotheses, and ``Test WER" means the average WER across the test sets of 9 accents.}
    \label{tab:tabel4}
    \centering
\begin{tabular}{ccc}
\toprule
\textbf{Speech Emb.} & \textbf{Hyp.}                                                           & \textbf{Test WER (\%)} \\ \midrule
\ding{52}                    & Word-level                                                              & \textbf{3.20}         \\
\ding{52}                    & Phoneme-level                                                           & 3.37         \\
\ding{52}                    & Word-level \& Phoneme-level & 3.46         \\ \bottomrule
\end{tabular}
\end{table}
\subsection{Ablation Study on Multi-modal GER}
Table~\ref{tab:tabel4} investigates the impact of different granularities of 1-best hypotheses with speech embeddings on multi-modal GER.
The combination of speech embeddings and 1-best word-level hypotheses achieves the best performance, as the fine-grained pronunciation information from speech and the coarse-grained semantic information from word-level hypotheses effectively enhance the LLMs' ability for accented speech transcription prediction.
The input of speech embeddings and 1-best phoneme-level hypotheses results in a performance decline, as the combination of fine-grained pronunciation information without semantic information fails to activate the LLMs' transcription prediction capabilities.
This phenomenon demonstrates the critical role of semantic information in enabling LLMs to predict transcriptions from hypotheses effectively.
However, the simultaneous input of speech embeddings, word-level hypotheses, and phoneme-level hypotheses into the LLMs further decreases performance.
This phenomenon can be attributed to the excessive redundancy and conflicts in the information, which overloaded the LLMs and impeded their transcription prediction capabilities.
Therefore, we find that the semantic information provided by word-level hypotheses is essential for GER, and excessive pronunciation information can hinder its performance.
Based on the conclusion, we can only propose multi-modal GER and multi-granularity GER separately, rather than integrating them into a single GER variant.
Multi-modal GER's limitation is its reliance solely on speech embeddings and 1-best word-level hypotheses, which constrains the LLMs' potential to fully exploit the semantic information contained within the N-best word-level hypotheses.

\subsection{Ablation Study on Dynamic Thresholds in HDMoLE}
Table~\ref{tab:tabel5} analyzes the effects of the dynamic threshold expert selection strategy in the HDMoLE.
When the dynamic threshold strategy is replaced with the static Top-K selection method, performance declines as the number of selected experts is reduced.
Employing a single dynamic threshold, whether global or local, equates to a Top-All expert selection strategy. 
While it achieves slightly better results than the static Top-All expert selection strategy without dynamic thresholds, its performance is still inadequate.
\begin{table}[h]
    \caption{Ablation study WER(\%) results for HDMoLE with different dynamic thresholds on our combined multi-accent English test sets. ``\#Threshold" denotes the number of dynamic thresholds in each HDMoLE layer, ``Type of Threshold" refers to local or global thresholds, ``Top-K" represents the top K experts to be selected, and ``Test WER" means the average WER across the test sets of 9 accents.}
    \label{tab:tabel5}
    \centering
\begin{tabular}{cccc}
\toprule
\textbf{\#Threshold} & \textbf{Type of Threshold} & \textbf{Top-K} & \textbf{Test WER (\%)} \\ \midrule
0                    & -                          & 1              & 3.24         \\ 
0                    & -                          & 3              & 3.23         \\
0                    & -                          & 6              & 3.19         \\
0                    & -                          & All            & 3.12         \\ \midrule
1                    & Local                      & -              & 3.07         \\
1                    & Global                     & -              & 3.09         \\
1                    & Local \& Global            & -              & 2.73         \\ \midrule
2                    & Local \& Global            & -              & \textbf{2.49}         \\ \bottomrule
\end{tabular}
\end{table}
\begin{figure}[h]
  \centering
  \includegraphics[width=1.0\linewidth]{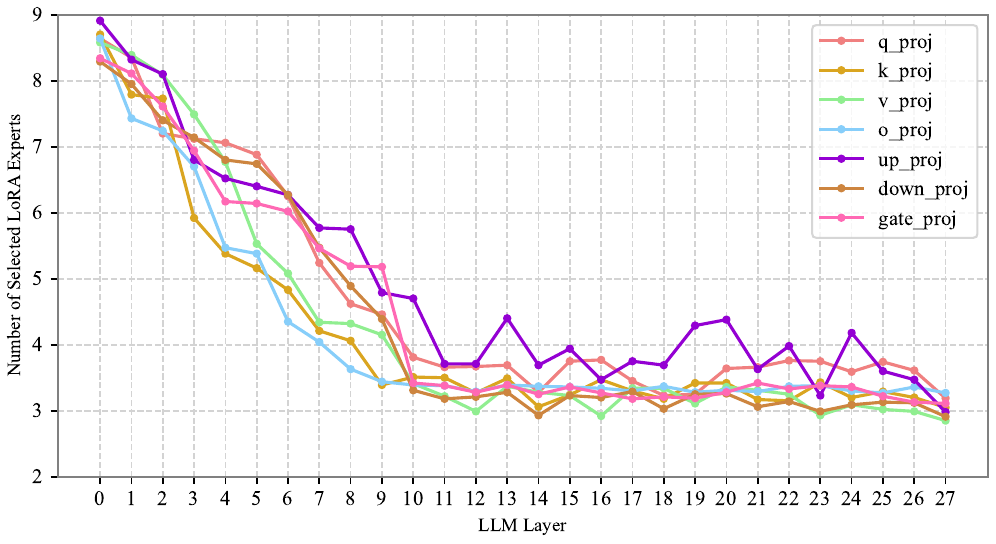}
  \caption{The average number of LoRA experts selected in the seven modules using HDMoLE within each LLM layer. The legend refers to seven modules where HDMoLE is employed.}
  \label{fig:fig5}
\end{figure}
This advancement is attributed to its ability to assign greater weights to beneficial LoRA experts in global or local weights, outperforming the static Top-All expert selection strategy.
Furthermore, by employing two distinct dynamic thresholds to independently regulate local and global weights, the model achieves more flexibility in expert selection.
Therefore, the dynamic thresholds strategy better accommodates the unique demands of each HDMoLE layer, enhancing overall HDMoLE performance.
Fig.~\ref{fig:fig5} visualizes the average number of selected LoRA experts across the seven modules within each LLM layer that employed HDMoLE.
A noticeable trend is observed where the number of selected LoRA experts diminishes across the first ten LLM layers and subsequently stabilizes.
This phenomenon indicates that the first ten layers of the LLM tend to leverage more comprehensive information, requiring HDMoLE to integrate a more significant number of LoRA experts. In contrast, the deeper layers of the LLM focus on more specific information, allowing HDMoLE to integrate fewer LoRA experts.
\begin{figure}[h]
    \centering
    \subfloat{
        \includegraphics[width=1.0\linewidth]{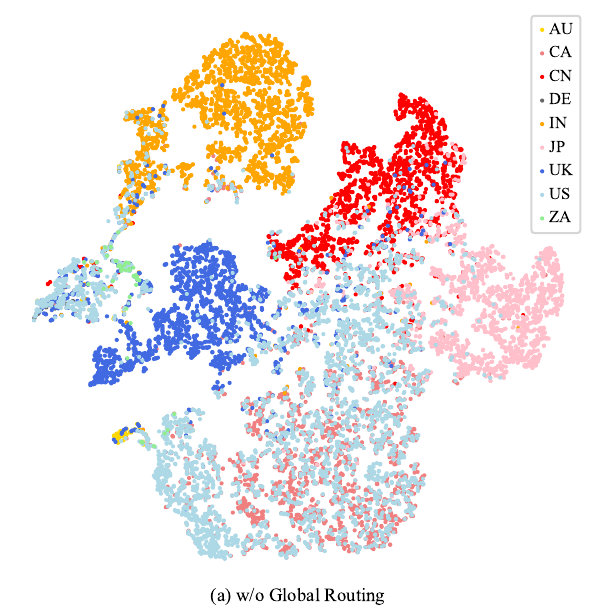}
        \label{fig:fig6}
    }\\
    \subfloat{
        \includegraphics[width=1.0\linewidth]{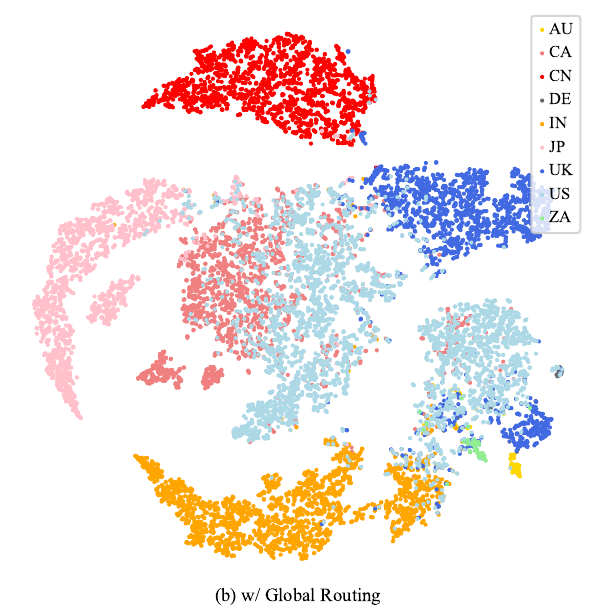}
        \label{fig:fig7}
    }
    \caption{T-SNE visualization of the local weights with or without global routing. The legend refers to different accents.}
    \label{fig:fig67}
\end{figure}
\begin{table}[h]
    \caption{Ablation study WER(\%) results for HDMoLE with different global routing on our combined multi-accent English test sets. ``AR ACC" refers to the accuracy of pre-trained AR models, and ``Test WER" means the average WER across the test sets of 9 accents.}
    \label{tab:tabel6}
    \centering
\begin{tabular}{l|ccccc}
\toprule
\textbf{AR ACC (\%)} & 0.00 & 50.79 & 71.29 & 83.77 & 90.51 \\ \midrule
\textbf{Test WER (\%)} & 3.08 & 3.15  & 2.89  & 2.67  & \textbf{2.49}  \\ \bottomrule
\end{tabular}
\end{table}
\subsection{Ablation Study on Hierarchical Routing in HDMoLE}
Table~\ref{tab:tabel6} examines the effects of hierarchical routing in HDMoLE.
When the AR ACC is 0, indicating the absence of global routing, the local routing still partially performs the task of combining LoRA experts.
At an AR ACC of 50.79\%, the global routing significantly disrupts the local routing, decreasing performance.
Once the AR ACC exceeds 71.29\%, global routing plays a significant role in guiding the local routing, leading to better LoRA expert combination outcomes.
Additionally, we visualize the local weights using t-SNE, as shown in Fig.~\ref{fig:fig67}.
The visualization results indicate that the global routing enhances the ability of the local routing to generate local weights that are more effective in distinguishing different accents.
Meanwhile, certain accents, such as American and Canadian accents, exhibit less distinct differentiation.
This phenomenon can be attributed to the geographical proximity of America and Canada, which results in a certain degree of similarity between their accents. Moreover, this similarity facilitates mutual improvement in speech recognition performance for American and Canadian accents.
Hierarchical routing takes advantage of the complementary performance resulting from this similarity, allowing it to combine the relevant LoRA experts for more accurate transcription predictions, especially for accents prone to confusion.
On the other hand, since HDMoLE keeps the LoRA experts frozen during training, they retain their exceptional ability to recognize corresponding accented speech. Hierarchical routing can assign significantly higher weights to their corresponding mono-accent LoRA experts for distinctly different accents to achieve accurate transcription predictions, such as Chinese and Japanese accents.
\begin{table}[h]
    \caption{Ablation study WER(\%) results for multi-granularity GER on our combined multi-accent English test sets. ``Word-level Hyp." means whether to use word-level hypotheses, ``Phoneme-level-Hyp." represents whether to use phoneme-level hypotheses, ``N-best" refers to the number of word-level or phoneme-level hypotheses, and ``Test WER" denotes the average WER across the test sets of 9 accents.}
    \label{tab:tabel7}
    \centering
\begin{tabular}{cccc}
\toprule
\textbf{Word-level Hyp.} & \textbf{Phoneme-level Hyp.} & \textbf{N-best} & \textbf{Test WER (\%)} \\ \midrule
\ding{52}                      & \ding{56}                         & 1               & 2.47         \\
\ding{52}                      & \ding{56}                         & 3               & 2.36         \\
\ding{52}                      & \ding{56}                         & 5               & 2.29         \\ \midrule
\ding{56}                      & \ding{52}                         & 1               & 2.49         \\
\ding{56}                      & \ding{52}                         & 3               & 2.44         \\
\ding{56}                      & \ding{52}                         & 5               & 2.41         \\ \midrule
\ding{52}                      & \ding{52}                         & 1               & 2.46         \\
\ding{52}                      & \ding{52}                         & 3               & 2.23         \\
\ding{52}                      & \ding{52}                         & 5               & \textbf{2.07}         \\ \bottomrule
\end{tabular}
\end{table}
\subsection{Ablation Study on Multi-granularity GER}
Table~\ref{tab:tabel7} evaluates how the granularity and the number of hypotheses affect multi-granularity GER.
The number of hypotheses also significantly impacts the error correction performance of the LLM.
The increase in the number of hypotheses correlates with improved error correction performance by the LLM, as more hypotheses increase the chances of including more accurate words or phonemes.
Moreover, the N-best word-level hypotheses outperform the N-best phoneme-level hypotheses (e.g., 1-best, 3-best, and 5-best), demonstrating that LLM is more adept at understanding the semantic information in word-level hypotheses, thereby facilitating error correction at the hypothesis level.
In comparison, the N-best phoneme-level hypotheses restrict the LLM to partial word-level error correction, offering a limited contribution to improving the LLM's error correction performance.
Optimal performance is achieved by concurrently leveraging word-level and phoneme-level hypotheses, where word-level hypotheses provide semantic information and phoneme-level hypotheses provide pronunciation information. This combination is more suitable for addressing accented speech recognition challenges.
Unfortunately, due to computational resource limitations, we cannot explore a greater number of hypotheses to substantiate our conclusions and investigate the boundary of hypothesis quantity for LLM error correction performance.
Furthermore, the failure to achieve significant enhancements with the use of only 1-best hypotheses in multi-granularity GER accentuates the inherent limitations of multi-modal GER where the use of 1-best hypotheses.
\section{Conclusions}
This study investigates more effective methods for improving GER to address the challenges posed by real-world accented speech scenarios.
We propose improved methods for GER based on the multi-granularity of accent speech recognition and the diversity of accents.
Firstly, we propose multi-modal GER by integrating pronunciation information from the speech modality and semantic information from the text modality. 
Subsequently, we implemented a three-stage training strategy for the multi-modal GER using accent-specific speech data to produce mono-accent LoRA experts.
We then integrate these LoRA experts in a single multi-modal GER model using our proposed HDMoLE method, which leverages hierarchical routing and dynamic thresholds, enabling the multi-modal GER model to address the challenges posed by diverse accents.
Finally, we propose the multi-granularity GER, which simultaneously leverages coarse-grained N-best word-level hypotheses and fine-grained N-best phoneme-level hypotheses. 
Multi-modal GER and multi-granularity GER leverage pronunciation and semantic information to achieve more accurate accented speech transcription predictions.
Moreover, HDMoLE can combine multiple mono-accent LoRA experts to address multi-accent scenarios.
We evaluate the effectiveness of our proposed methods using a multi-accent English dataset. Compared to the vanilla whisper-large-v3 baseline model, our methods achieve a relative WER reduction of 67.35\%.
Ablation study on multi-modal GER, HDMoLE, and multi-granularity GER confirm their theoretical validity and practical feasibility.
In future work, we plan to explore applying our proposed methods to larger foundational models and extending them to more complex acoustic conditions in real-world scenarios.
\bibliographystyle{IEEEbib}
\bibliography{refs}

\end{document}